\documentclass[journal]{IEEEtran}
\usepackage{cite}

\ifCLASSINFOpdf
\usepackage[pdftex]{graphicx}
\DeclareGraphicsExtensions{.pdf,.jpeg,.png}
\else
\usepackage[dvips]{graphicx}
\DeclareGraphicsExtensions{.eps}
\fi
\usepackage{epsfig,epsf,epstopdf,graphicx}
\usepackage[cmex10]{amsmath}
\usepackage{amssymb}
\usepackage{amsthm }
\usepackage{nicefrac}
\usepackage{hyperref}
\usepackage{xcolor}
\hypersetup{colorlinks=true}
\usepackage{tabularx}

\theoremstyle{theorem}

\theoremstyle{definition}

\theoremstyle{plain}

\theoremstyle{plain}

\newcommand\Tstrut{\rule{0pt}{2.6ex}}         
\newcommand{\bb}{{\textbf{b}}}

\usepackage{algorithm}
\usepackage{multirow}
\usepackage{algcompatible}
\usepackage[bottom]{footmisc}

\PassOptionsToPackage{dvipsnames}{xcolor}
\usepackage{tikz}
\usetikzlibrary{calc}

\ifCLASSOPTIONcompsoc
  \usepackage[caption=false,font=normalsize,labelfont=sf,textfont=sf]{subfig}
\else
  \usepackage[caption=false,font=footnotesize]{subfig}
\fi
\usepackage{fixltx2e}
\usepackage{afterpage}
\usepackage{float}
\usepackage{stfloats}

\makeatletter

\begin{document}

%

\title{Composite Generalized Quadratic Noise Modulation via Signal Addition: Towards Higher Dimensional Noise Modulations}

\author{Hadi~Zayyani, \IEEEmembership{Member,~IEEE,} Mohammad Salman, \IEEEmembership{Senior Member,~IEEE},
Felipe A. P. de Figueiredo, Rausley A. A. de Souza, \IEEEmembership{Senior Member,~IEEE}


\thanks{This work has been partially funded by the xGMobile Project (XGM-AFCCT-2024-9-1-1) with resources from EMBRAPII/MCTI (Grant 052/2023 PPI IoT/Manufatura 4.0), by CNPq (302085/2025-4, 306199/2025-4), and FAPEMIG (APQ-03162-24).}
\thanks{H.~Zayyani is with the Department
of Electrical and Computer Engineering, Qom University of Technology (QUT), Qom, Iran (e-mails: zayyani@qut.ac.ir).}
\thanks{M.~Salman is with College of Engineering and Technology, American University of the Middle East, Egaila, 54200, Kuwait (e-mail: mohammad.salman@aum.edu.kw).}
\thanks{F. A. P. de Figueiredo, and R. A. A. de Souza are with the National Institute of Telecommunications (Inatel), Santa Rita do Sapucaí, Brazil. (e-mail: felipe.figueiredo@inatel.br, rausley@inatel.br).}



\vspace{-0.8cm}}


\maketitle
\thispagestyle{plain}
\pagestyle{plain}

\begin{abstract}
This letter proposes superposing two Generalized Quadratic Noise Modulators (GQNM) by simply adding their outputs. It creates a 16-ary noise modulator that resembles QAM modulators in classical communication. It modulates the information bits on four different means and four different variances. It could also be applied to reach higher-order modulations than 16-ary schemes by adding the outputs of more than two modulators, which is not discussed in detail in this letter and left for future work. By selecting the parameters necessary for satisfying the theoretical distinguishability conditions provided in the paper, we can reach better performances in comparison to the Kirchhoff-Law Johnson Noise (KLJN) modulator and the GQNM modulator, which is verified by the simulations. The better result in terms of smaller Bit Error Probability (BEP) is achieved by increasing the complexity in the modulator, the transmitter, and the detectors in the receiver.
\end{abstract}


%
\IEEEpeerreviewmaketitle

\section{Introduction}
\label{sec:Intro}

\IEEEPARstart{N}{oise} modulation is a new modulation scheme in the newly arisen noise communication framework. Noise communication systems utilize the inherent random thermal noise of resistors to convey the information bits with zero power consumption. In addition to the power consumption advantage, the noise communication systems are also secure. The roots of noise communication date back to 2005, in the pioneering work of Kish \cite{Kish05}. The noise communication framework is well studied in the following works by Kish \cite{Kish06}-\cite{Kish10}. In \cite{Kish05}, Kish proposed the concept of a zero-power stealthy communication system from an applied physics perspective, utilizing two resistors with different impedances.

In a sequel, the same author proposed a Kirchhoff-Law-Johnson Noise (KLJN) secure key exchange scheme, in which the laws of physics provide the basis for unconditionally secure communication: Kirchhoff's law and thermal noise of two pairs of resistors \cite{Kish06}. Moreover, Kish et al. proposed totally secure classical networks with multipoint telecloning (teleportation) of classical bits through loops with Johnson-like noise \cite{Kish06_1}. In addition, \cite{Kish10} discusses the effect of wire resistance on the noise voltage and current. Also, some other advanced concepts of noise communication are discussed in \cite{Ming08}-\cite{Kape22}. These concepts are the basis for a communication scheme stretching from two to two thousand kilometers \cite{Ming08}. The first attempt to calculate the Bit Error Probability (BEP) of the KLJN noise communication scheme was presented in \cite{Saez13}, with further reduction of BEP explored in \cite{Saez13_1}. A more advanced method for calculating the BEP of the KLJN communication system was introduced in \cite{Smul14}, while the investigation of noise properties in a KLJN communication system was conducted in \cite{Ging14}. A generalized KLJN approach was proposed in \cite{Vadai15}, and security and performance analyses of both the KLJN and generalized KLJN secure key exchange protocols were provided in \cite{Ming15} and \cite{Ming17}, respectively. Finally, an experimental realization of an ultralow-power wireless communication system by selectively connecting or disconnecting an impedance-matched resistor and an antenna was demonstrated in \cite{Kape22}.



Although the contributions mentioned above originated largely in non-communication fields such as physics, a breakthrough was made by Basar in 2023 \cite{Basar23}. This work introduced the concept of noise communication to the communications engineering community by formalizing the problem from a rigorous signal processing perspective. \cite{Basar23} formulated a new framework for the BEP calculation of the KLJN noise communication system and suggested two novel detectors that further reduce the BEP. Moreover, the same author presented a noise modulation scheme for wireless communication in which the noise samples are directly fed to the antenna system \cite{Basar24}. Furthermore, a binary on-off keying for noise modulation is presented in \cite{Anjos25}. In addition, an innovative joint energy harvesting and communication scheme is suggested for future Internet-of-Things (IoT) devices by leveraging the emerging noise modulation technique \cite{Yapici25}.

Recently, in \cite{Tasci25}, a Flip-KLJN secure noise communication is proposed where a pre-agreed intermediate level, such as high/low (H/L), triggers a flip of the bit map value during the bit exchange period. More recently, a KLJN-based thermal noise modulation has been proposed, which represents a viable solution for secure IoT communication at ultra-low power levels \cite{Salem25}. More recently, a Generalized Quadratic Noise Modulator (GQNM) has been suggested in which non-Gaussian noise is utilized for the waveforms, and two low and high voltages are used to have non-zero mean waveforms \cite{ZayyArxiv25}. Additionally, a resistor hopping scheme for wire-secure noise communication is suggested in \cite{ZayyArxivRH25}.


This paper develops higher-order noise modulations beyond the quadratic scheme proposed in \cite{ZayyArxiv25}. We achieve this by superposing two or more GQNM outputs through simple addition. This paper provides a detailed explanation of how to add two GQNM modulators. The generalization to more than two GQNM modulators is feasible but challenging to implement and is left for future work. The superposing scheme results in a 16-ary noise modulator which resembles the QAM modulations in classical communication. The 16-ary Composite GQNM (CGQNM) modulator has four different levels in the mean dimension and has four different levels in the variance dimension. Moreover, the detectors of four bits in CGQNM are designed, and some simple thresholds are suggested for the detectors, which are applicable when the distinguishability condition is satisfied. This paper theoretically establishes the distinguishability conditions and demonstrates that parameters can be selected to meet them. Simulations confirm the superior performance of the proposed CGQNM modulator, which significantly outperforms both the classical KLJN and single GQNM modulators.


The paper is organized as follows. Section~\ref{sec:ProblemForm} discusses the system model and preliminaries. In Section~\ref{sec: prop}, the proposed superposing generalized noise modulation framework is presented and developed. Theoretical distinguishability conditions are provided in section~\ref{sec: ana}. Simulation results are presented in Section~\ref{sec: Simulation}, while conclusions are drawn in Section~\ref{sec: con}.

\section{System Model and Preliminaries}\label{sec:ProblemForm}

In this section, we briefly present the concept of noise modulation as introduced in \cite{Basar24} and generalized noise modulation introduced in \cite{ZayyArxiv25}. In a noise modulator, a single information bit is used to select a random Gaussian noise source, for instance, from a low resistor $R_L$ or a high resistor $R_H$. For a bit '0', a low-variance Gaussian noise is transmitted, whereas for a bit '1', a high-variance noise is transmitted. The variance is treated as a distinguishable dimension (or degree of freedom). The noise samples, comprising $N$ samples per bit duration, are fed to a wireless antenna system via a baseband or even an IQ-modulator \cite{Basar24}. At the receiver, the baseband samples are processed for bit detection. This is typically achieved through threshold-based detection by estimating the sample variance to recover the information embedded in the noise variance. This classical noise modulation scheme is analogous to a binary communication system. Subsequently, a generalized quadratic noise modulation (GQNM) scheme is proposed in \cite{ZayyArxiv25}. In the generalized scheme, which results in a quadratic modulation, two measures of separability are utilized. The first one is the variance, such as used in a classical noise modulator, which is controlled by a sub-bit that switches between low and high resistors. We call this sub-bit the variance sub-bit. The second is the mean of the noise, which is distinguishable when we add the voltage biases of $m_L$ or $m_H$ to the KLJN of resistors based on the additional second sub-bit. We nominate that sub-bit as the mean sub-bit. The block diagram of a GQNM is shown in Fig.~1. In the next section, we propose a low-complex method to superpose generalized noise modulators.

\begin{figure}[h]
\begin{center}
\includegraphics[scale=0.46]{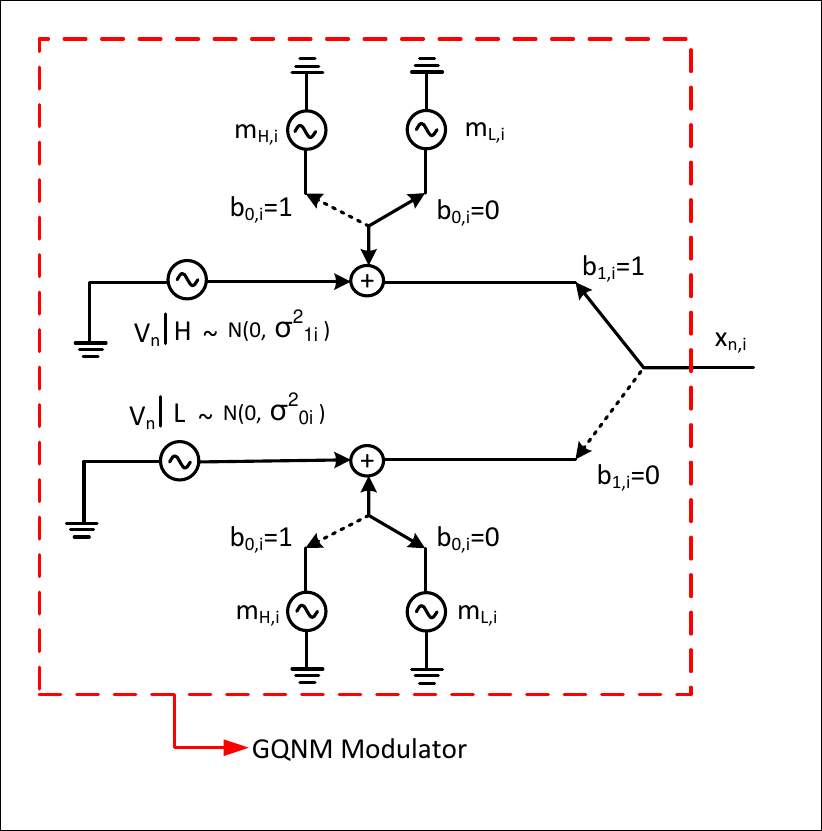}
\end{center}
\vspace{-0.5 cm}
\caption{Block diagram of the generalized quadratic noise modulation scheme \cite{ZayyArxiv25}.}
\label{fig1}
\vspace{-0.3cm}
\end{figure}
\begin{figure}[h]
\begin{center}
\includegraphics[scale=0.4]{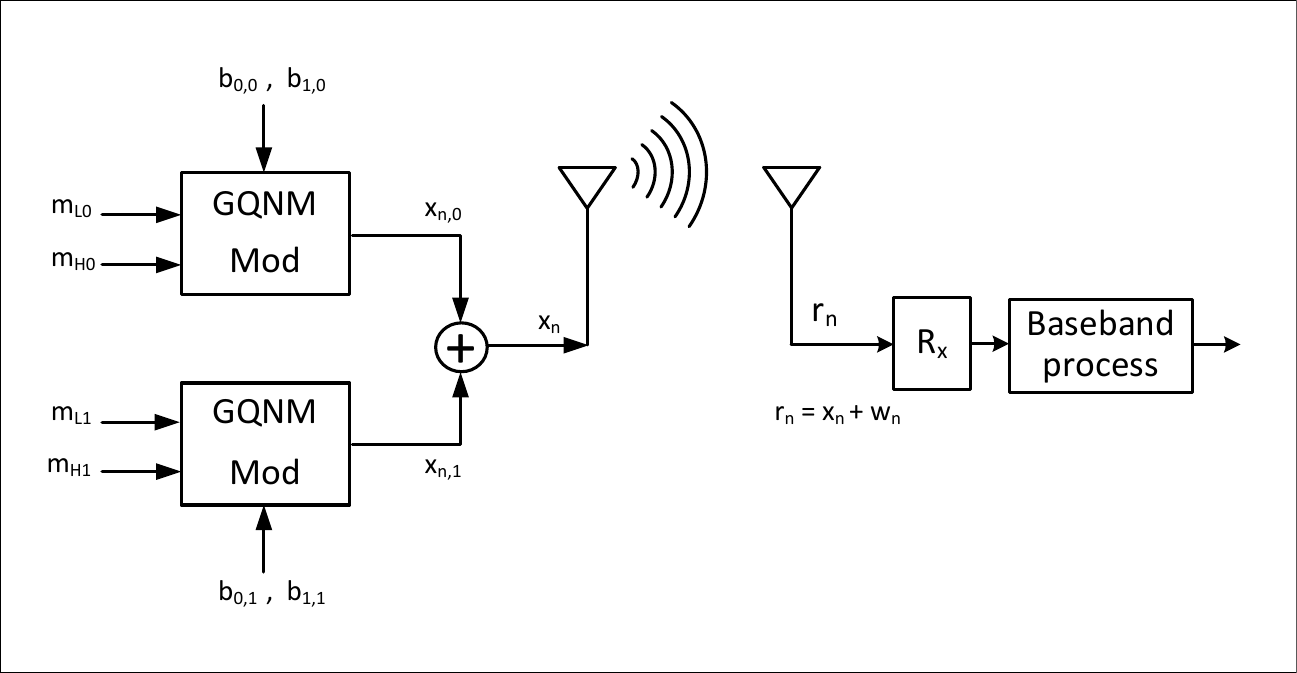}
\end{center}
\vspace{-0.5 cm}
\caption{Superposition of two GQNM modulators by adding their signals, resulting in a 16-ary composite noise modulation scheme.}
\label{fig2}
\vspace{-0.3cm}
\end{figure}

\section{Superposing Generalized Noise Modulators}
\label{sec: prop}

In this section, we propose superposing the GQNM modulators by simply adding them. We develop the superposition by adding only two GQNM modulators, but the rationale can be generalized to more than two GQNM modulators. We show that by adding two GQNM modulators, we reach a 16-ary composite noise modulator (CNM). Similarly, by adding $s\ge 2$ number of GQNM modulators, we can reach a $2^{2s}$-ary CNM modulator as well. We focus on the case of $s=2$ for analytical tractability, leaving the generalization to $s>2$ for future work. The block diagram of the overall noise communication system formed by superposing two ($s=2$) GQNM modulators is depicted in Fig.~2.
In this figure, we have the transmitted noise symbol from each modulator being given by
\begin{align}
x_{n,i}=m_{n,i}+v_{n,i}\quad i=0,1,
\end{align}
where
\begin{align}
m_{n,i}=\left\{
          \begin{array}{ll}
            m_{L_i}, & b_{0,i}=0 \\
            m_{H_i}, & b_{0,i}=1,
          \end{array}
        \right.
\end{align}
and
\begin{align}
v_{n,i}\sim\left\{
             \begin{array}{ll}
               N(0,\sigma^2_{0i}), & b_{1,i}=0 \\
               N(0,\sigma^2_{1i}), & b_{1,i}=1,
             \end{array}
           \right.
\end{align}
where $b_{0,i}$, and $b_{1,i}$ are the bits of the $i$'th sub-channel, and $m_{L_i}$, $m_{H_i}$ with assumption of $m_{L_i}<m_{H_i}$ are the bias voltages of the $i$'th sub-channel, and $\sigma^2_{0i}$, $\sigma^2_{1i}$ are the variances of low resistor and high resistors of the $i$'th sub-channel with the assumption of $\sigma_{0i}<\sigma_{1i}$. Therefore, we have four bias voltages of $m_{L_0}$, $m_{L_1}$, $m_{H_0}$, and $m_{H_1}$ and four noise variances of $\sigma^2_{00}$, $\sigma^2_{01}$, $\sigma^2_{10}$, and $\sigma^2_{11}$. For better modeling, similar to the classical KLJN noise communication system, we assume $m_{H_i}=\alpha m_{L_i}$, and $m_{L_1}=\beta m_{L_0}$ with $\alpha, \beta>1$ and $\alpha>\beta$. Typically, we consider $\alpha=50$ and $\beta=5$ in practice. A similar assumption is done for the four different variances of $\sigma^2_{00}$, $\sigma^2_{01}$, $\sigma^2_{10}$, and $\sigma^2_{11}$, which are $\sigma^2_{1i}=\eta\sigma^2_{0i}$ and $\sigma^2_{i1}=\gamma\sigma^2_{i0}$ with $\eta, \gamma>1$ with $\gamma>\eta$. The typical values we used in our simulations are $\gamma=40$ and $\eta=10$. Thus, the overall independent parameters are $m_{L_0}$, $\alpha$, $\beta$, $\sigma^2_{00}$, $\eta$, and $\gamma$.

\begin{figure}[h]
\begin{center}
\includegraphics[scale=0.52]{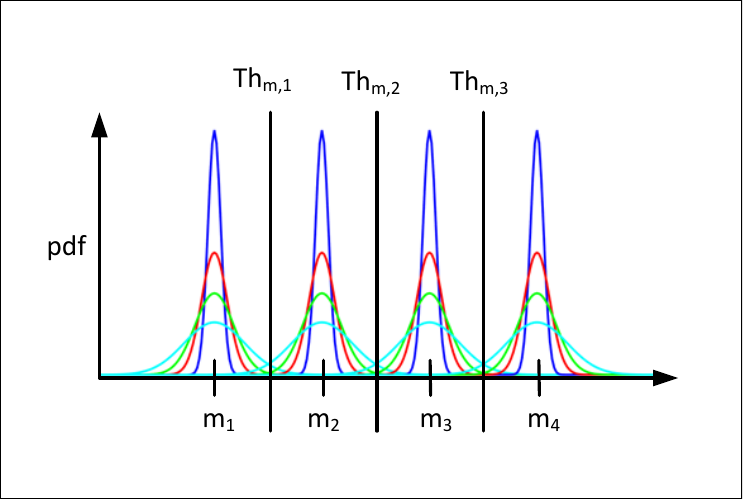}
\end{center}
\vspace{-0.5 cm}
\caption{The PDF of transmitted noise symbol, $x_n$.}
\label{fig3}
\end{figure}

\begin{table}[!b]
\caption{Mean and variance of Gaussian random variable, $x_n$, in the 16 different cases.}
\centering
 \begin{tabular}{p{16mm}||p{13mm}|p{13mm}|p{13mm}|p{13mm}} \hline
 \Tstrut $(b_{0,1},b_{1,1})$ & (00) & (01)& (10) & (11) \\\hline \hline

 \Tstrut $(b_{0,0},b_{1,0})=(00)$
 & $\!\!\! \begin{aligned} m^{(n)}=m_1\\ \sigma^{2,(n)}=\sigma^2_1 \end{aligned} $	
 & $\!\!\!\begin{aligned} m^{(n)}=m_1\\ \sigma^{2,(n)}=\sigma^2_3 \end{aligned}$	
 & $\!\!\!\begin{aligned} m^{(n)}=m_3\\ \sigma^{2,(n)}=\sigma^2_1 \end{aligned}$	
 &	$\!\!\!\begin{aligned} m^{(n)}=m_3\\ \sigma^{2,(n)}=\sigma^2_3 \end{aligned}$  \\ \hline

 \Tstrut $(b_{0,0},b_{1,0})=(01)$
 & $\!\!\! \begin{aligned} m^{(n)}=m_1\\ \sigma^{2,(n)}=\sigma^2_2 \end{aligned} $	
 & $\!\!\!\begin{aligned} m^{(n)}=m_1\\ \sigma^{2,(n)}=\sigma^2_4 \end{aligned}$	
 & $\!\!\!\begin{aligned} m^{(n)}=m_3\\ \sigma^{2,(n)}=\sigma^2_2 \end{aligned}$	
 &	$\!\!\!\begin{aligned} m^{(n)}=m_3\\ \sigma^{2,(n)}=\sigma^2_4 \end{aligned}$  \\ \hline


 \Tstrut $(b_{0,0},b_{1,0})=(10)$
 & $\!\!\! \begin{aligned} m^{(n)}=m_2\\ \sigma^{2,(n)}=\sigma^2_1 \end{aligned} $	
 & $\!\!\!\begin{aligned} m^{(n)}=m_2\\ \sigma^{2,(n)}=\sigma^2_3 \end{aligned}$	
 & $\!\!\!\begin{aligned} m^{(n)}=m_4\\ \sigma^{2,(n)}=\sigma^2_1 \end{aligned}$	
 &	$\!\!\!\begin{aligned} m^{(n)}=m_4\\ \sigma^{2,(n)}=\sigma^2_3 \end{aligned}$  \\ \hline


 \Tstrut $(b_{0,0},b_{1,0})=(11)$
 & $\!\!\! \begin{aligned} m^{(n)}=m_2\\ \sigma^{2,(n)}=\sigma^2_2 \end{aligned} $	
 & $\!\!\!\begin{aligned} m^{(n)}=m_2\\ \sigma^{2,(n)}=\sigma^2_4 \end{aligned}$	
 & $\!\!\!\begin{aligned} m^{(n)}=m_4\\ \sigma^{2,(n)}=\sigma^2_2 \end{aligned}$	
 &	$\!\!\!\begin{aligned} m^{(n)}=m_4\\ \sigma^{2,(n)}=\sigma^2_4 \end{aligned}$  \\ \hline


\end{tabular}
 \begin{tabular} {l}
\\
$m_1\triangleq m_{L_0}+m_{L_1}=(\beta+1)m_{L_0}$, $m_2\triangleq m_{H_0}+m_{L_1}=(\beta+\alpha)m_{L_0}$,\\ $m_3\triangleq m_{L_0}+m_{H_1}=(1+\alpha\beta)m_{L_0}$, $m_4\triangleq m_{H_0}+m_{H_1}=\alpha(\beta+1)m_{L_0}$\\
$\sigma^2_1\triangleq \sigma^2_{00}+\sigma^2_{01}=(1+\gamma)\sigma^2_{00}$, $\sigma^2_2\triangleq \sigma^2_{10}+\sigma^2_{01}=(\eta+\gamma)\sigma^2_{00}$,\\ $\sigma^2_3\triangleq \sigma^2_{00}+\sigma^2_{1}=(1+\gamma\eta)\sigma^2_{00}$, $\sigma^2_4\triangleq \sigma^2_{10}+\sigma^2_{11}=\gamma(1+\eta)\sigma^2_{00}$
\end{tabular}
\label{Table_1}
\end{table}

The transmitted noise sample is assumed to be
\begin{align}
x_n=x_{n,0}+x_{n,1}=m^{(n)}+v^{(n)},
\end{align}
where $m^{(n)}=m_{n,0}+m_{n,1}$ and $v^{(n)}=v_{n,0}+v_{n,1}$ is the zero-mean Gaussian with variance $\sigma^{2,(n)}$. By adding the outputs of these two modulators, we have a symbol which consists of a four bit vector of $\bb_n=[\bb^T_{n,0},\bb^T_{n,1}]^T$, where $\bb_{n,i}=[b_{0,i} b_{1,i}]^T$ is the $i$'th sub-channel bit vector, which consists of two bits. In the case of $s=2$, we have $2s=4$ bits in a symbol duration of modulation. It results in a $2^{2s}=16$-ary composite noise modulator. Based on these four bits, we have 16 states in the CNM. In each of these 16 cases, we have a transmitted noise symbol $x_n$, which is Gaussian with mean $m_n$ and variance $\sigma^2_n$. All the 16 cases are shown in Table~1, along with the mean and variance of each case. As we see from the table, the 16 states are in 16 different states, where there is no exact match in both the mean and variance. Indeed, the random Gaussian variable $x_n\sim N(m^{(n)},\sigma^{2,(n)})$ is a mixture of 16 Gaussian random variables, and the Probability Density Function (PDF) of a typical occurrence is depicted in Fig.~3. In fact, following the definition of $m_j$ and $\sigma^2_j$ for $1\le j\le 4$ in Table~1, we straightforwardly have $m_1<m_2<m_3<m_4$ and $\sigma^2_1<\sigma^2_2<\sigma^2_3<\sigma^2_4$, which is in accordance to what is represented in Fig.~3. As can be seen from Fig.~3, if the means of $m_j$ are separated enough with respect to their variances, the four categories of the Gaussian distribution can be separated or distinguished by threshold detectors, with the suitable thresholds shown in Fig.~3. If the distinction in the mean is satisfied (which is discussed in the next section), the bits related to the mean dimension can be detected very simply by comparing the sample mean of the total samples in a symbol duration with a threshold. Therefore, we have
\begin{align}
(\hat{b}_{0,0},\hat{b}_{0,1})=\left\{
                                \begin{array}{ll}
                                  (1,0), & \hat{m}_n<\mathrm{Th}_{m,1} \\
                                  (1,0), &  \mathrm{Th}_{m,1}<\hat{m}_n<\mathrm{Th}_{m,2}\\
                                  (0,1), & \mathrm{Th}_{m,2}<\hat{m}_n<\mathrm{Th}_{m,3} \\
                                  (1,1), & \hat{m}_n>\mathrm{Th}_{m,3},
                                \end{array}
                              \right.
\end{align}
where $\hat{m}_n=\frac{1}{N}\sum_{k}x_{n,k}$ in which $N$ is the total number of samples in a symbol duration, and $1\le k\le N$ is the index of noise samples. For far distinguished Gaussian distributions, the thresholds are selected as $\mathrm{Th}_{m,r}=\frac{m_r+m_{r+1}}{2}$ for $1\le r\le 3$.
After separating the means and detecting the two sub-bits of the mean dimension, we should detect two other sub-bits that are related to the variance dimension. In this regard, we calculate the sample variance as
\begin{align}
\hat{\sigma}^{2,(n)}=\frac{1}{N}\sum_{k=1}^N(x_{n,k}-\mathrm{E}(x_{n,k}))^2,
\end{align}
where $\mathrm{E}\{x_{n,k}\}=\hat{m}_n=\frac{1}{N}\sum_{k}x_{n,k}$. The sample variance is itself a random variable with $\mathrm{E}\{\hat{\sigma}^{2,(n)}|f\}=\frac{1}{N}\sum_{k=1}^N\mathrm{Var}(x_{n,k}|f)=\sigma^2_f$, where $f$ is the index of Gaussian category in the mean dimension with $f=1,2,3,4$. As we discussed more in the next section, the sample variance has a mixture of Gaussian (MoG) distributions with means equal to $\sigma^2_f, \forall f$. The PDF of the sample variance is depicted in Fig.~4 for a typical case. If the separability or distinguishability in the variance dimension holds (which is discussed in the next section), the sub-bits of the variance dimension are detected as
\begin{align}
(\hat{b}_{1,0},\hat{b}_{1,1})=\left\{
                                \begin{array}{ll}
                                  (0,0), & \hat{\sigma}^{2,(n)}<\mathrm{Th}_{v,1} \\
                                  (1,0), &  \mathrm{Th}_{v,1}<\hat{\sigma}^{2,(n)}<\mathrm{Th}_{v,2}\\
                                  (0,1), & \mathrm{Th}_{v,2}<\hat{\sigma}^{2,(n)}<\mathrm{Th}_{v,3} \\
                                  (1,1), & \hat{\sigma}^{2,(n)}>\mathrm{Th}_{v,3},
                                \end{array}
                              \right.
\end{align}
where the thresholds are simply selected as $\mathrm{Th}_{v,r}=\frac{\sigma^2_r+\sigma^2_{r+1}}{2}$ for $1\le r\le 3$.

\section{Distinguishability of a mixture of Gaussian distributions}
\label{sec: ana}
In this section, we analyze the required conditions for the sharp distinguishability of MoG components. In the simulations, we show that we can design the CNM modulator to achieve these conditions.

To investigate the conditions of sharp distinguishability\footnote{This means that the Gaussian distributions are far apart from each other such that they are only partially overlapped.} of Gaussian components in the mean dimension, the mean and variances of the sample mean should be calculated. We have $\hat{m}^{(n)}\sim N(m_i,\sigma^2_i)$ in which the means $m_i$ and variances $\sigma^2_i$ are calculated in Table.~1. From Fig.~3, the sufficient distinguishability conditions for the mean dimension are
\begin{align}
\label{eq: mcond}
m_r+3\sigma_4\ll m_{r+1}-3\sigma_4,\quad 1\le r\le 3,
\end{align}
where, since the means of $m_1$ and $m_2$ are nearer to each other than any other successive pairs of means, all three conditions in (\ref{eq: mcond}) reflect a single condition of $m_2-m_1\gg6\sigma_4$, which is simplified to
\begin{align}
\label{eq: mcondf}
\alpha m_{L_0}\gg6\sqrt{\gamma(1+\eta)}\sigma_{00}.
\end{align}

Since we have six independent free parameters as described before, the condition of (\ref{eq: mcondf}) is satisfied easily by either selecting a high value for $m_{L_0}$ or selecting a small value for $\sigma_{00}$ for design purposes.

\begin{figure}[h]
\begin{center}
\includegraphics[scale=0.52]{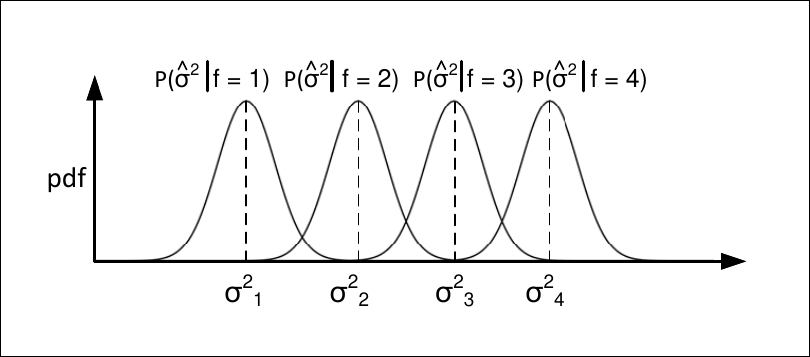}
\end{center}
\vspace{-0.5 cm}
\caption{The PDF of sample variance.}
\label{fig3}
\end{figure}

To investigate the condition of sharp distinguishability of Gaussian components in the variance dimension, the mean and variance of the sample variance should be calculated. The mean of the sample covariance is calculated in Section~\ref{sec: prop} and is equal to
\begin{align}
m_f\triangleq \mathrm{E}\{\hat{\sigma}^{2,(n)}|f\}=\sigma^2_f,
\end{align}
where $f$ is the index of the sub-Gaussian category in the mean dimension.

Using Central Limit Theory (CLT), the sample variance also has a Gaussian distribution with mean $m_f$ and variance $\mathrm{Var}(\hat{\sigma}^{2}|f)$. To calculate $\mathrm{Var}(\hat{\sigma}^{2}|f)$, we have
\begin{align}
\label{eq: varff}
\mathrm{Var}(\hat{\sigma}^{2}|f)=\mathrm{E}\{\hat{\sigma}^4|f\}-\mathrm{E}^2\{\hat{\sigma}^2|f\}=\mathrm{E}\{\hat{\sigma}^4|f\}-\sigma^4_f,
\end{align}
where the superscript $(n)$ is removed for the sake of simplicity. To calculate $\mathrm{E}\{\hat{\sigma}^4|f\}$ in (\ref{eq: varff}),  we note that we have
\begin{align}
\label{eq: mmsig}
\mathrm{E}\{\hat{\sigma}^4|f\}&=\mathrm{E}\{\Big(\frac{1}{N}\sum_{k=1}^N(x_{n,k}-\mathrm{E}\{x_{n,k}\})^2\Big)^4\}\nonumber\\
&=(\frac{1}{N})^4\sigma^2_f\mathrm{E}\{e^4|f\},
\end{align}
where $e=\sum_{k=1}^N(\frac{x_{n,k}-\mathrm{E}\{x_{n,k}\}}{\sigma_f})^2$ is a chi-square distributed with $(N-1)$ degree of freedom, i.e., $e\sim\chi^2_{N-1}$. Thus, knowing that the $m$'th order moment of a $X\sim\chi^2_k$ is equal to $\mathrm{E}\{X^m\}=2^m\frac{\Gamma(\frac{k}{2}+m)}{\Gamma(\frac{k}{2})}$, we have
\begin{align}
\label{eq: m4f}
m_{4,e}=\mathrm{E}\{e^4\}=16\frac{\Gamma(\frac{N-1}{2}+4)}{\Gamma(\frac{N-1}{2})}=16\frac{\Gamma(\frac{N+7}{2})}{\Gamma(\frac{N-1}{2})},
\end{align}
where it is independent of the index $f$. Putting (\ref{eq: m4f}) into (\ref{eq: mmsig}) and then into (\ref{eq: varff}), we have
\begin{align}
\label{eq: varff1}
\sigma^2_{\mathrm{var},f}\triangleq\mathrm{Var}(\hat{\sigma}^{2}|f)=16(\frac{1}{N})^4\sigma^2_f\frac{\Gamma(\frac{N+7}{2})}{\Gamma(\frac{N-1}{2})}-\sigma^4_f.
\end{align}

Following Fig.~4, the sufficient distinguishability conditions in the variance domain are
\begin{align}
\label{eq: varcond}
\sigma^2_f+3\sigma_{\mathrm{var},f}\ll\sigma^2_{f+1}-3\sigma_{\mathrm{var},f+1}, \quad 1\le f\le 3.
\end{align}

Assuming that the most stringent condition among those in (\ref{eq: varcond}) corresponds to $f=1$, due to the closest proximity of the means, the final sufficient distinguishability condition simplifies to
\begin{align}
\label{eq: varcondf}
\sigma^2_2-\sigma^2_1\gg 3\sigma_{\mathrm{var},1}+3\sigma_{\mathrm{var},2}, \quad 1\le f\le 3,
\end{align}
which is further simplified to the final condition of
\begin{align}
\label{eq: varcondff}
\eta\sigma^2_{00}\gg &3\sigma_1\sqrt{16(\frac{1}{N})^4\frac{\Gamma(\frac{N+7}{2})}{\Gamma(\frac{N-1}{2})}-\sigma^2_1}+\nonumber\\
&3\sigma_2\sqrt{16(\frac{1}{N})^4\frac{\Gamma(\frac{N+7}{2})}{\Gamma(\frac{N-1}{2})}-\sigma^2_2}.
\end{align}

\begin{figure}[tb]
\begin{center}
\includegraphics[width=8cm]{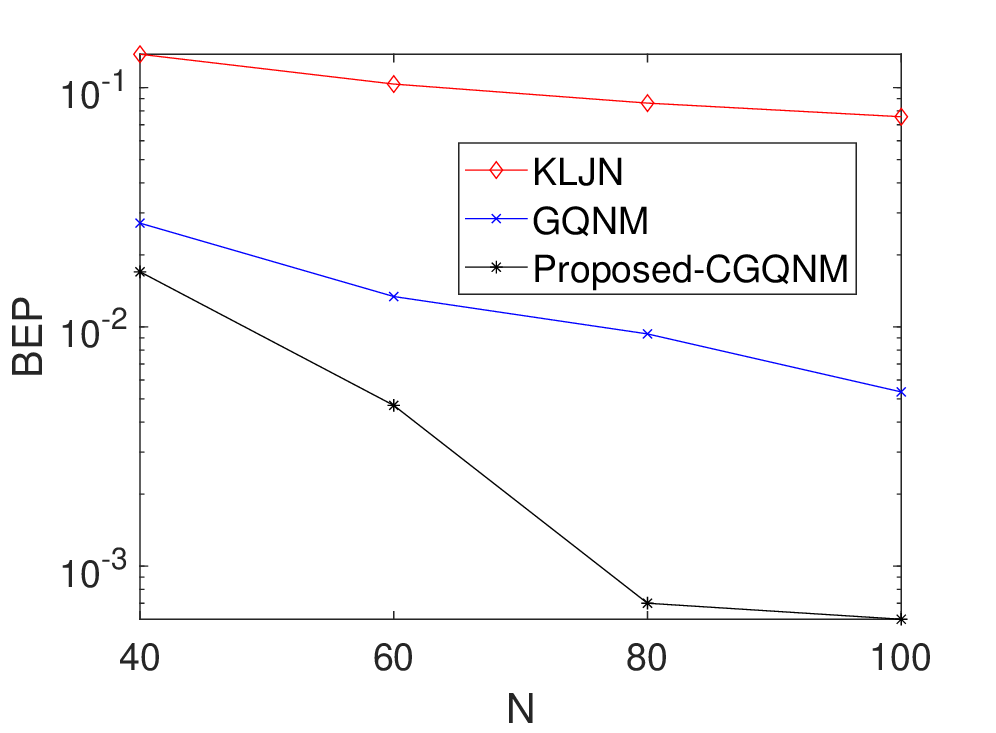}
\end{center}
\vspace{-0.5 cm}
\caption{Simulated BEPs versus $N$.}
\label{fig1}
\vspace{-0.3cm}
\end{figure}

\section{Simulation Results}
\label{sec: Simulation}
In this section, the performance of the proposed CGQNM is investigated. The first GQNM modulator has the parameters of $\sigma_{00}=1e-5$, $\sigma_{10}=\sqrt{\gamma}\sigma_{00}=1.4142e-5$, $m_{L_0}=1e-3$, and $m_{H_0}=\alpha m_{L_0}=20e-3$ with $\alpha=20$, $\gamma=20$. For the second GQNM modulator, we select $\eta=5$ and $\beta=5$ and hence we have $\sigma_{01}=2.2361e-5$, $\sigma_{11}=1e-4$, $m_{L_1}=5e-3$, and $m_{H_1}=0.1$. These selected parameters are selected such that the distinguishability conditions in (\ref{eq: mcondf}) and (\ref{eq: varcondff}) are satisfied. The random bits are generated uniformly. Three noise modulators are simulated: the KLJN modulator \cite{Basar24}, the single GQNM modulator with a Gaussian waveform \cite{ZayyArxiv25}, and the proposed CGQNM. To fairly compare the algorithms, we use the same sampling rate for all three schemes, i.e., they have the same number of samples per bit duration, which is assumed to be $N=100$, if not otherwise stated. The performance metric is the BEP. The channel noise is assumed to be Gaussian with zero mean and standard deviation of $\sigma_w=2e-5$, if not otherwise stated. Two experiments were performed.

The first experiment investigated the effect of the number of samples per bit duration, or $N$, which determines the sampling rate. In this case, the value of $N$ is varied between 40 and 100. The BEP versus $N$ is depicted in Fig.~\ref{fig1}. It shows that the proposed CGQNM has a lower BEP than single GQNM and KLJN modulators, and as $N$ increases, the BEP decreases and performance improves.

In the second experiment, the effect of the standard deviation of channel noise $\sigma_w$ is investigated. The $\sigma_w$ is varied from $1e-5$ to $5e-5$, where this range is the range of change in BEP from low to high. The BEP versus the $\sigma_w$ is shown in Fig.~\ref{fig2}. It demonstrates again that the proposed CGQNM is the best algorithm, and the BEP decreases when $\sigma_w$ decreases. The better results of CGQNM are achieved by increasing the complexity of the modulator on the transmitter side and the detectors on the receiver side.

\begin{figure}[tb]
\begin{center}
\includegraphics[width=8cm]{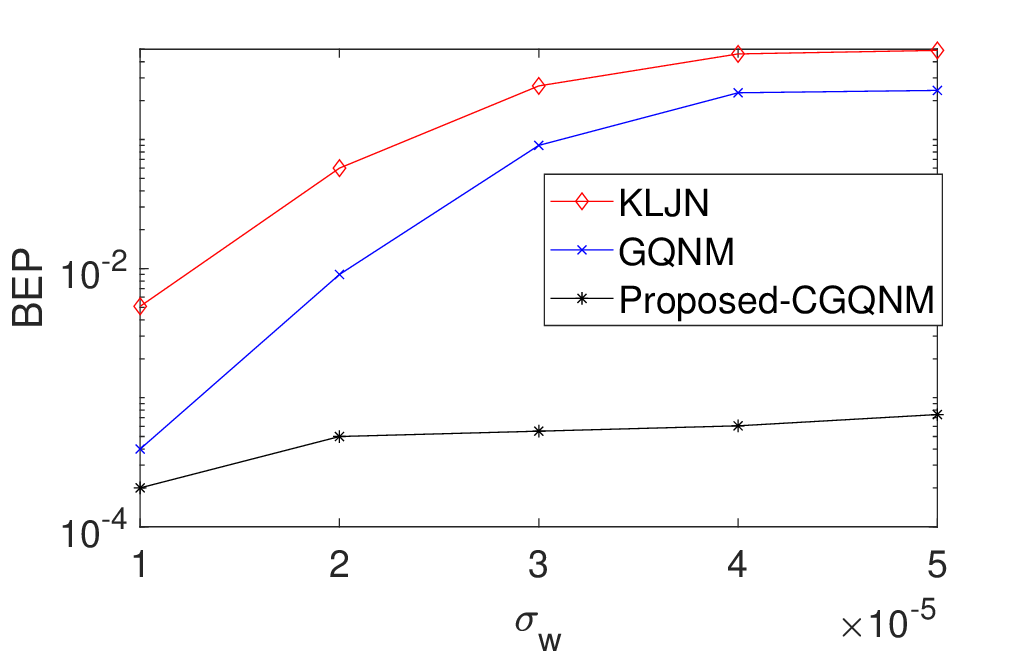}
\end{center}
\vspace{-0.5 cm}
\caption{Simulated BEPs versus standard deviation of noise $\sigma_w$.}
\label{fig2}
\vspace{-0.3cm}
\end{figure}

\section{Conclusion and future work}
\label{sec: con}

This letter proposes increasing the modulation order of quadratic noise modulation further to 16-ary modulation or even more. The details of how to develop a 16-ary noise modulation scheme based on simply adding two GQNM modulators are provided. In fact, similar to QAM modulations in classical communication, which have two I-Q dimensions, here, in noise modulation, we have two dimensions of mean and variance. We can modulate the information bits on both the mean and the variance. Using one GQNM modulator results in a quadratic modulation, while adding the outputs of two GQNM modulators results in a 16-ary modulation. The detectors' details of the CGQNM modulator satisfying the distinguishability conditions of different Gaussian distributions are derived in the letter. Simulation results show the advantage of less BEP of the proposed CGQNM modulator than the KLJN noise modulator and the GQNM modulator. It is accomplished by increasing the complexity of the modulator and detectors in both sides of the transmitter and receiver. Future work includes generalizing the 16-ary noise modulation to $2^{2s}$-ary higher order modulations in a unified formulation, where $s$ is the number of GQNM modulators that participate in the superposition. Other future work is to calculate the BEP of the proposed higher-order modulations in a closed form, similar to what was done for the GQNM modulator in \cite{ZayyArxiv25}.

\end{document}